\documentclass[fleqn,usenatbib]{mnras}

\usepackage{graphicx}
\usepackage{times}
\usepackage{grffile} %
\usepackage{newtxmath}
\usepackage{bm}
\usepackage{cleveref}
\usepackage{xspace}
\usepackage{booktabs}
\usepackage[flushleft]{threeparttable}
\usepackage{siunitx}
\usepackage[super]{nth}
\DeclareSIUnit{\hred}{\mathit{h}}
\DeclareSIUnit{\Msun}{M_\odot}
\DeclareSIUnit\pc{pc}
\DeclareSIUnit\kpc{kpc}
\DeclareSIUnit\Mpc{Mpc}
\DeclareSIUnit\lightyear{ly}
\usepackage{xcolor}
\usepackage{soul}
\usepackage[normalem]{ulem}
\usepackage{enumerate}
\usepackage{tabularx}

\definecolor{bubbles}{rgb}{0.91, 1.0, 1.0}
\definecolor{aquamarine}{rgb}{0.5, 1.0, 0.83}
\definecolor{bubblegum}{rgb}{0.99, 0.76, 0.8}
\definecolor{bluebell}{rgb}{0.74, 0.74, 0.92}
\definecolor{dollarbill}{rgb}{0.72, 0.93, 0.6}

\crefname{table}{Table}{Tables}
\Crefname{table}{Table}{Tables}

\crefname{equation}{Eq.}{Eqs.}
\Crefname{equation}{Equation}{Equations}

\crefname{figure}{Fig.}{Figs.}
\Crefname{figure}{Figure}{Figures}

\crefname{section}{Section}{Sections}
\Crefname{section}{Section}{Sections}
\crefname{appendix}{Appendix}{Appendices}
\Crefname{appendix}{Appendix}{Appendices}

\newcommand{\ie}{i.e.\xspace}

\newcommand{\Cov}{C}

\newcommand{\ramses}{\textsc{Ramses}\xspace}

\newcommand{\genetic}{\textsc{genetIC}\xspace}
\newcommand{\adaptahop}{\textsc{AdaptaHOP}\xspace}
\newcommand{\tangos}{\textsc{Tangos}\xspace}

\renewcommand{\vec}[1]{\bm{#1}}
\newcommand{\mat}[1]{\mathsf{#1}}

\title[Environment, halo mass and concentration]{%
    The causal effect of environment on halo mass and concentration
}

\author[C.~Cadiou et al.]{Corentin~Cadiou$^{1}$\thanks{c.cadiou@ucl.ac.uk},
Andrew~Pontzen$^{1}$,
Hiranya~V.~Peiris$^{1,2}$,
Luisa~Lucie-Smith$^{3}$
\\
$^{1}$Department of Physics and Astronomy, University College London, Gower Street, London WC1E 6BT, United-Kingdom\\
$^{2}$The Oskar Klein Centre for Cosmoparticle Physics, Department of Physics, Stockholm University, AlbaNova, Stockholm SE-106 91, Sweden\\
$^{3}$Max Planck Institute for Astrophysics, Karl-Schwarzschild-Strasse 1, 85748 Garching, Germany
}

\date{Accepted 2021 September 10. Received 2021 September 9; in original form 2021 July 7.}
\pubyear{2021}

\begin{document}
\label{firstpage}
\pagerange{\pageref{firstpage}--\pageref{lastpage}}
\maketitle

\begin{abstract}
    Understanding the impact of environment on the formation and evolution of dark matter halos and galaxies is a crucial open problem. Studying statistical correlations in large simulated populations sheds some light on these impacts, but the causal effect of an environment on individual objects is harder to pinpoint. Addressing this, we present a new method for resimulating a single dark matter halo in multiple large-scale environments.
    In the initial conditions, we `splice' (\ie{} insert) the Lagrangian region of a halo into different Gaussian random fields, while enforcing consistency with the statistical properties of $\Lambda$CDM\@.
    Applying this technique, we demonstrate that the mass of halos is primarily determined by the density structure inside their Lagrangian patches, while the halos' concentration is more strongly affected by environment. The splicing approach will also allow us to study, for example, the impact of the cosmic web on accretion processes and galaxy quenching.
\end{abstract}


\section{Introduction}

The growth of dark matter halos and galaxies can be most accurately computed using numerical simulations. Understanding the physical origin of environmental quenching \citep[e.g.][]{kauffmann_EnvironmentalDependenceRelations_2004,peng_MassEnvironmentDrivers_2010}, intrinsic alignments \citep[e.g.][]{tempel_EvidenceSpinAlignment_2013,chisari_IntrinsicAlignmentsGalaxies_2015} or colour gradients in the cosmic web \citep{laigle_COSMOS2015PhotometricRedshifts_2018,kraljic_GalaxyEvolutionMetric_2018} are some of the most fundamental open problems in galaxy formation.
However, attaining a physical understanding of these effects of cosmological environment on individual galaxies is complicated by the wide variety of possible configurations that are generated by the Gaussian random initial conditions (ICs).

Currently, the main approach to disentangling the impact of environmental factors on galaxy formation is statistical in nature \citep{aubert_OriginImplicationsDark_2004,danovich_CoplanarStreamsPancakes_2012,codis_ConnectingCosmicWeb_2012,kraljic_GalaxiesFlowingOriented_2019,martizzi_BaryonsCosmicWeb_2020}. Analytic models can provide hypotheses for the causal relationships between ICs and final halos \citep[e.g.][]{press_FormationGalaxiesClusters_1974,sheth_EllipsoidalCollapseImproved_2001,hahn_TidalEffectsEnvironment_2009,codis_SpinAlignmentsCosmic_2015,musso_HowDoesCosmic_2018} but it is difficult to test these hypotheses at the level of individual halos \citep{borzyszkowski_ZOMGHowCosmic_2017,LucieSmith19}.

In this work, we extend the `genetic modification' (GM) technique \citep{roth_GeneticallyModifiedhaloes_2016}, which is designed specifically to construct controlled experiments in cosmological galaxy and halo formation. Previously, GM has been used to control the mass, merger history \citep{Pontzen17,rey_QuadraticGeneticModifications_2018} and angular momentum \citep{cadiou_AngularMomentumEvolution_2021} of individual objects. Our extension aims to manipulate instead the large-scale environment, while leaving the density structure of a target object's Lagrangian patch untouched.

We extend the code \genetic{} \citep{stopyra_GenetICNewInitial_2020}, to embed the ICs that will eventually collapse into a halo into new environments. This can be seen as a `gene-splicing' operation, combining two Gaussian random fields into a single realisation. We apply this technique to investigate how the mass and concentration of halos in dark matter simulations are affected by environment.

The paper is structured as follows: we first present qualitatively the gene-splicing method and the set of numerical simulations used throughout the paper in \cref{sec:qualitative-presentation-splicing}.
We then present their analysis in \cref{sec:results}.
Finally, we summarise and discuss our findings in \cref{sec:discussion-conclusion}.
A more detailed mathematical derivation of the gene-splicing method can be found in \cref{sec:splicing-mathematical-derivation}.

\section{Methods}
\label{sec:qualitative-presentation-splicing}

\begin{figure}
    \includegraphics[width=\columnwidth]{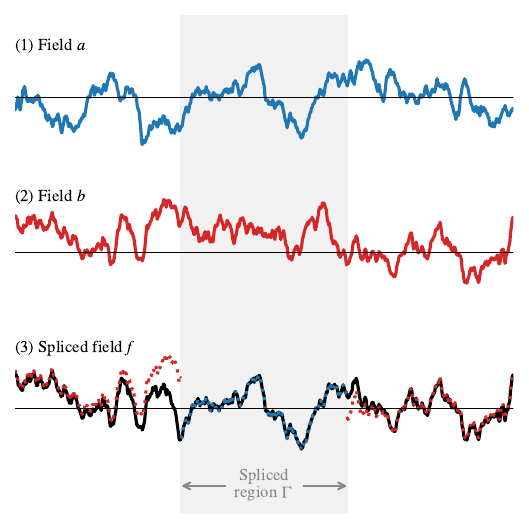}
    \caption{
        Illustration of the gene-splicing procedure applied to one dimensional initial conditions. We draw a field $a$ (in blue) and another independent field $b$ (in red).
        We obtain the new initial conditions (in black) by `splicing' a given region of $a$ into $b$.
        The spliced field has the value of $a$ in the spliced region and rapidly converges to the value of $b$ outside it, while remaining maximally consistent with the Gaussian random field statistics.
    }\label{fig:splicing-1D}
\end{figure}

In this section, we first present the `gene-splicing' technique;
a more formal derivation can be found in \cref{sec:splicing-mathematical-derivation}. We will then discuss how it has been applied to produce a suite of simulations for this first study.

The splicing operation is applied to the linear initial conditions, which we generate at $z=100$. We start from two Gaussian random fields representing the overdensity of independent realisations, denoted $a$ and $b$, and select an arbitrary region $\Gamma$.
To obtain the results in this paper, we choose $\Gamma$ to be the Lagrangian region of a $z=0$ halo (i.e.\ the region that its constituent particles occupied at $z=100$).

The splicing operation finds a new field $f$ which satisfies $f(x)=a(x)$ inside $\Gamma$, but which closely approximates $b(x)$ elsewhere in the simulation volume. It is not possible  to set $f(x)=b(x)$ outside $\Gamma$ because this would cause discontinuities on the boundary; such discontinuities are incompatible with the assumption of a Gaussian random field. Instead, we minimise the $\chi^2$ of the field difference $f(x) - b(x)$. This approach has been motivated at length by~\cite{roth_GeneticallyModifiedhaloes_2016} and~\cite{rey_QuadraticGeneticModifications_2018}, and leads to fields that are maximally likely in the Gaussian random ensemble under the constraints.
Given the spliced density, we then use the Zel’dovich approximation to generate a corresponding set of particles with new positions and velocities.
These are used as initial conditions for a new $N$-body simulation.

The algorithm described above is equivalent to altering the field $b(x)$ with a list of modifications specifying the new value of $f(x)$ at every point $x_i$ in $\Gamma$. However, applying the existing GM algorithm to this problem becomes quickly impractical as the number of points in $\Gamma$ increases, requiring $\mathcal{O}\left(N^d \times N_\mathrm{pt}\right)$ memory, where $N$ is the number of cells in each direction, $d$ is the number of dimensions and $N_\mathrm{pt}$ is the number of constrained points.
To circumvent this problem, we instead solve the $\chi^2$ difference minimisation iteratively using a gradient descent method.
We have implemented the method within the code \genetic{} \citep{stopyra_GenetICNewInitial_2020} in v1.3 \citep{andrew_pontzen_2021_5079937}.
Further details can be found in \cref{sec:splicing-mathematical-derivation}.

A 1D example of a spliced Gaussian random field is illustrated in \cref{fig:splicing-1D}; the splicing region $\Gamma$ is indicated by grey shading. The independent fields $a$ and $b$ are shown in the top two panels; the spliced field $f$ is shown in the bottom panel (solid line) along with the relevant portions of the original fields for comparison (dotted lines). The spliced field $f$ can be seen to obey our requirements: it traces $a$ perfectly inside $\Gamma$; is continuous on the boundary of $\Gamma$; and closely approximates $b$ at large distances from $\Gamma$. The rate at which $f$ converges to $b$ depends both on the correlation function (or equivalently the power spectrum) and on the difference between fields $a$ and $b$ around the splicing region boundary. In this test, the reduced $\chi^2$ of realisations $a$, $b$ and $f$ are $1.00$, $1.03$ and $0.99$ respectively (with \num[group-separator={,},group-minimum-digits=3]{1499} degrees of freedom), indicating that $f$ is a likely draw from the underlying distribution despite being constructed artificially.

\begin{figure}
    \centering

    \includegraphics[width=\columnwidth]{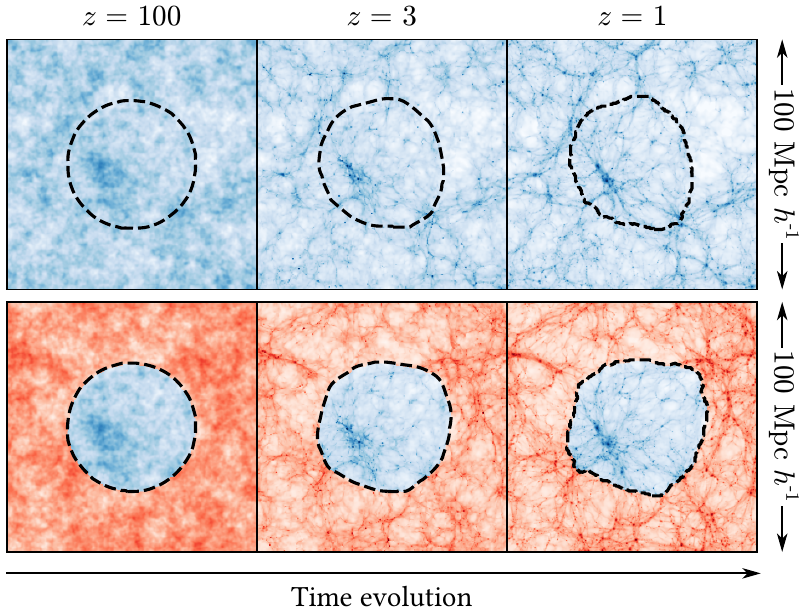}

    \caption{Slices of the dark matter density field evolved from an unmodified set of ICs (top row) and corresponding spliced ICs (bottom row).
        Regions evolving from the original ICs are coloured in blue; the new external region is coloured in red.
        The sphere (dashed lines) is tidally distorted over cosmic time, leading to differences between the two simulations in terms of the shape of the boundary. 
        Structures in the spliced region (bottom row, in blue) can be mapped on to their counterparts in the original simulation (top row).
        Conversely, outside this region, the matter density fields in the two simulations bear no resemblance to each other.
    }\label{fig:illustration_splicing}
\end{figure}

Having shown how splicing works in a 1D example, we next illustrate in \cref{fig:illustration_splicing} the cosmological evolution of a 3D spliced field.
The top left panel shows our reference ICs at redshift $z=100$; we use a $256^3$ grid in a domain of size \SI{100}{Mpc\per\hred} for a mass resolution of $M_\mathrm{DM} = \SI{7.7e9}{\Msun}$.
The transfer function is computed using \textsc{Camb} \citep{lewis_EfficientComputationCMB_2000} and cosmological parameters consistent with the values of~\cite{planckcollaboration_Planck2018Results_2018}.
The initial conditions are then evolved using \ramses{} \citep{teyssier_CosmologicalHydrodynamicsAdaptive_2002a}, as illustrated in the top row. Gravity is solved using a particle-mesh approach on an adaptive mesh.
We allow the mesh to be refined wherever it contains more than 8 dark matter particles.
The effective minimal force resolution reached by the simulation is \SI{9}{kpc} physical.

Next, we select a region $\Gamma$ in the ICs of the reference simulation.
As an illustrative example, in \cref{fig:illustration_splicing} we splice a sphere of comoving radius \SI{25}{Mpc\per\hred}.
Finally, we draw an independent overdensity field, and splice the sphere into it to form the new ICs; the result is shown in the bottom left panel. The region which is identical to the original ICs is shown in blue, while the external region is shown in red.
We evolve the new initial conditions using an identical simulation configuration to the original.

\begin{figure}
    \includegraphics[width=\columnwidth]{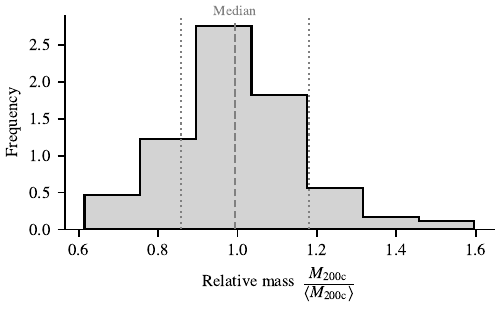}
    \caption{When splicing halos into a new realisation, their mass changes due to environmental effects. For our six halos, each simulated in ten different environments,  we find that the change in mass is modest. The histogram shows the new mass divided by the mean over the ten realisations. Vertical lines indicate the median (dashed) and \SI{68}{\percent} credible interval (dotted), showing that the mass typically scatters only by $\pm\SI{15}{\percent}$.
    }%
    \label{fig:mass_ratio_distribution}
\end{figure}

The time evolution of the sphere in the reference (top row) and  spliced (bottom row) simulations  can now be compared.
We indicate the edge of the sphere (dashed black line), defined by the set of particles that it contains in the ICs as a function of time.
The edge of the region is deformed by non-linear structure formation, becoming less spherical with time.
This deformation depends on the long-range tidal effect of the region outside the sphere and so the shape of the patches increasingly differs between the two simulations.

The density field  within the sphere is identical, by construction, in the two sets of ICs. The subsequent interior gravitational evolution is similar; but it has small differences, due to the differing large-scale gravitational forces. The impact of these changes on halos is the focus of this paper.
By contrast, far from the sphere, the ICs are unrelated between the two simulations, and structures in one simulation cannot be mapped to the other. In the case illustrated, a large cosmic void is present in the rightmost region of the unaltered simulation, while a massive filament forms in the spliced simulation.

In the remainder of the paper, we will study how the large-scale environment contributes to setting the mass and concentration of dark matter halos, as an example of the gene-splicing technique's promise.
For this purpose, we performed a reference simulation with identical cosmological and numerical parameters to the example described above, in a domain of size \SI{50}{Mpc\per\hred}, for a mass resolution of $M_\mathrm{DM} = \SI{9.7e8}{\Msun}$ and an effective minimal force resolution of \SI{2}{kpc} physical.
From this unmodified simulation, we selected six dark matter halos with masses between\footnote{The individual masses are $\{3.2, 3.3, 5.3, 5.9, 7.2, 8.6\} \times 10^{13}\,\mathrm{M}_{\odot}$.} $10^{13}$ and $10^{14}\,\mathrm{M_\odot}$ at $z=0$.
We select all their member particles as computed by the halo finder~-- including those in any of their subhalos~-- and trace these back to the ICs to obtain the Lagrangian patch.
At this point, we have six patches that will eventually form a dark matter halo in the reference simulation. We separately spliced each of these six patches into $10$ independent realisations of the box, for a total of $60$ new ICs which were evolved to $z=0$.

We extract halo catalogues using \adaptahop{} \citep{aubert_OriginImplicationsDark_2004} and the parameters presented in~\cite{tweed_BuildingMergerTrees_2009a} with the `Most massive Substructure Method' and a minimum number of \num{200} particles per halo.
We analyse the catalogues using \tangos{} \citep{pontzen_TANGOSAgileNumerical_2018}, which we employ to extract the virial radius $R_\mathrm{200c}$, virial mass $M_{200c}$ and concentration parameter $c$ as we will describe below.

\begin{figure}
    \includegraphics[width=\columnwidth]{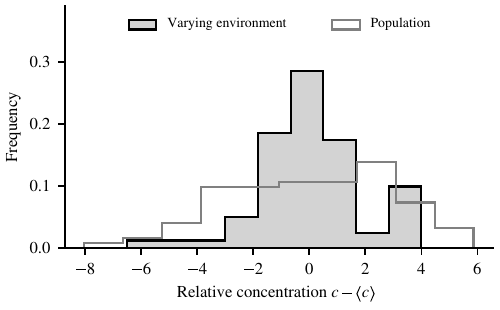}
    \caption{
        The scatter in the concentration induced by placing halos in a new environment is highly significant. For each halo, we calculate the scatter around its mean concentration in the ten environments. The shaded histogram shows the resulting distribution for all six halos, which can be compared to the scatter in concentration within the population (light histogram).
        At least half the scatter of the concentration can be attributed to the effect of environment.
    }%
    \label{fig:concentration_diff_distribution}
\end{figure}

\begin{figure*}
    \includegraphics*[width=\textwidth]{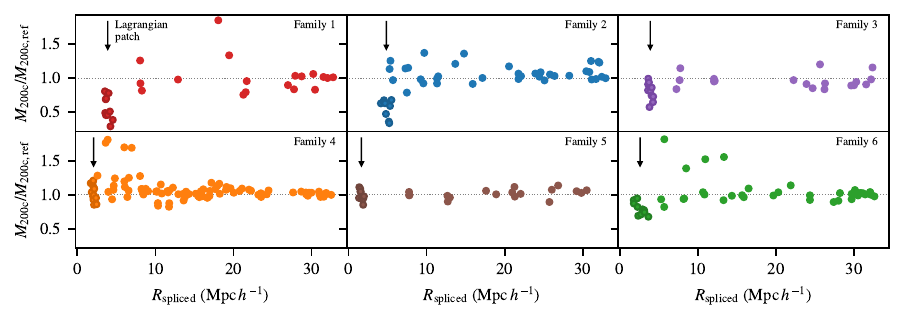}
    \caption{
        The ratio of the virial mass $M_\mathrm{200c}$ of the spliced halos to the reference halo, for the six reference halos.
        We highlight the simulations where the spliced region includes only the Lagrangian patch (darker symbols) and their mean $R_\mathrm{spliced}$ (black arrow); all other simulations use a splicing that has been expanded.
        The mass converges to the reference mass with increasing size of the spliced region at $z=0$, $R_\mathrm{spliced}$.
    }\label{fig:expansion-vs-mass}
\end{figure*}
\section{Results}%
\label{sec:results}

We now investigate the effect of environment on dark matter halos' masses. Our set of sixty simulations corresponds to ten environmental realisations around each of six central halos.
For each of the six halos, we compute the mean virial mass $\langle M_{\mathrm{200c}} \rangle$ over the ten realisations. We then calculate, for each realisation, the ratio of its mass to this mean:
\begin{equation}
    r = \frac{M_\mathrm{200c}}{\langle M_\mathrm{200c} \rangle}.
\end{equation}
This yields $60$ measurements of $r$, which are plotted as a histogram in \cref{fig:mass_ratio_distribution}; the masses are scattered by $\pm\SI{15}{\percent}$ around the halo's mean value.

Next, as an example of a more detailed structural property of halos, we measure the concentration parameter using the approach presented by \citet[][]{klypin_MultiDarkSimulationsStory_2016}; see their Equations~(18)-(20).
The NFW concentration parameter, $c$, is estimated using the implicit solution to
\begin{align}
    \frac{V^2_\mathrm{circ,max}}{V_\mathrm{200c}^2} &= \frac{c}{x_\mathrm{max}} \frac{f(x_\mathrm{max})}{f(c)},\\
    f(x) &\equiv \ln(1+x) - \frac{1}{1+x},\\
    x_\mathrm{max}& = 2.163.
\end{align}
Here $V^2_\mathrm{circ}(r) = GM(<r)/r$ is the circular velocity, $V_\mathrm{circ,max}$ is its maximum value for $0\leq r \leq R_\mathrm{200c}$ and $V_\mathrm{200c} = V_\mathrm{circ}(R_\mathrm{200c})$.
We measure the circular velocities in \num{100} logarithmically spaced radial bins between $R_\mathrm{200c}/100$ and $R_\mathrm{200c}$. We use this procedure because it is much more stable than fitting the NFW profile directly through $\chi^2$ optimisation, which suffers from significant degeneracies. We verified the numerical stability of the \citet[][]{klypin_MultiDarkSimulationsStory_2016} estimator by calculating the change in $c$ for all our halos between two adjacent timesteps, finding that its r.m.s.\ variation is only $\pm 10\%$. This is negligible compared to the population scatter that we will discuss below.

For each of the six halo families, we compute $\langle c \rangle$, where the average is taken over the ten environments. We then calculate a distribution of $c - \langle c \rangle$ over all sixty simulations. To
contextualise this distribution, we create a second ensemble, consisting of all 88 halos in the original reference run in the same mass window as the six reference halos, $10^{13} < M_{\mathrm{200c}}/M_{\odot} < 10^{14}$.
We then calculate $c - \langle c \rangle$ over this entire second population.
The difference in the statistics of these two ensembles captures the effect of the environment.

The results are shown in \cref{fig:concentration_diff_distribution}.
The two distributions are non-Gaussian; in order to compare them quantitatively, we compute the \SI{68}{\percent} and \SI{90}{\percent} credible intervals.
The \SI{68}{\percent} interval for the spliced distribution (shaded histogram), characterising the impact of varying environment alone, is $[-1.0,1.8]$. By contrast, the corresponding credible interval of the concentration of the entire population is $[-3.2,2.7]$.
 When using \SI{90}{\percent} credible intervals, the ranges expand to $[-2.1, 3.9]$ (spliced population) and $[-4.5, 4.0]$ (entire population).
Therefore, between half and $70\%$ of the scatter in the concentration at fixed mass can be attributed to the effect of environment.

Having looked at the effect of splicing the Lagrangian patch of halos into new environments, we now consider splicing larger regions. As discussed in \cref{sec:qualitative-presentation-splicing}, the size and shape of the spliced region can be chosen arbitrarily. Physically, one would expect that as the size of the spliced region expands, the influence of the external environment must become negligible because of the finite correlation length in $\Lambda$CDM\@. Accordingly, we expect the variation between environments of any measured halo property to become small.

We performed an additional set of 211 simulations, using three outer realisations around the same six inner families, but expanding the spliced region to progressively include all matter within some distance from the Lagrangian patch. As the region is expanded, it becomes progressively more spherical. We quantify the size of the resulting spliced regions at $z=0$ by an effective radius $R_{\mathrm{spliced}}$, where
\begin{equation}
    R_\mathrm{spliced}^3 = \frac{3}{4\pi} \frac{M_\mathrm{region}}{\langle \rho \rangle_\mathrm{v}},
\end{equation}
where $M_{\mathrm{region}}$ is the total mass in the spliced region, and $\langle \rho \rangle_{\mathrm{v}}$ is the volume-weighted mean density in the region at $z=0$.

The results are shown in \cref{fig:expansion-vs-mass}. Each panel uses simulations from one of our six families, showing how the final halo mass divided by the reference (unspliced) halo mass changes as the patch is expanded. Qualitatively, the halo mass converges towards the reference value as the splice radius becomes larger, as expected. This agrees with the work of \citet{LucieSmith19} who found that the information relevant to determining the mass of halos is localised within scales that are somewhat larger than their Lagrangian patches. However, we caution that a quantitative measure of the convergence radius using our method would require a considerably larger box size than the \SI{50}{Mpc\per\hred} used in the present study.

\section{Discussion and conclusions}%
\label{sec:discussion-conclusion}

We have presented `gene-splicing', a method for resimulating a chosen halo within a variety of environments, while respecting the statistical properties of $\Lambda$CDM initial conditions. This is an extension of the `genetic modification' approach \citep{roth_GeneticallyModifiedhaloes_2016,rey_QuadraticGeneticModifications_2018}, in which controlled experiments are carried out on a target halo, while the environment is minimally changed.

Manipulating Gaussian random fields in order to obtain insight into structure formation is an increasingly important tool \citep{aragon-calvo_MIPEnsembleSimulation_2016,Pontzen16InvertedICs,sawala_SettingStageStructures_2021b}.  Because structure formation is localised, it is often desirable to make modifications in real space. This, however, requires a careful treatment to maintain consistency with $\Lambda$CDM correlation structure. Our approach to doing so follows in a long tradition of solving linear constrained systems in cosmological contexts
\citep{bertschinger_PathIntegralMethods_1987,hoffman_ConstrainedRealizationsGaussian_1991,1996MNRAS.281...84V,bond_HowFilamentsGalaxies_1996,2006ApJ...637L..93R,elsner_EfficientWienerFiltering_2013}.

As a first demonstration of the gene-splicing method, we showed that at least half the scatter in the mass-concentration relation can be attributed to the effect of the large-scale environment.
This complements the results of~\cite{roth_GeneticallyModifiedhaloes_2016}, where it was shown that the time of collapse (encapsulated by the local density field) is not able on its own to account for the scatter in this relation.  We also showed that as the size of the spliced patch increases, the variation in mass
decays towards zero in accordance with physical expectations. However, due to running a large number of simulations (274) we used a relatively small box of \SI{50}{Mpc\per\hred}. With the splicing approach, larger boxes would be needed to robustly measure the size of the region which contains information about halo collapse.

While the focus of this paper was on mass and concentration of dark matter halos, many properties of halos and galaxies are affected by  their environment, and in future work, we will explore the underlying causal connections. For example, there is an observed correlation between galaxy quenched fraction and closeness to the nearest cosmological filament \citep{laigle_COSMOS2015PhotometricRedshifts_2018,kraljic_GalaxiesFlowingOriented_2019}, whose causal origin is as yet unclear
\citep{romano-diaz_ZOMGIIDoes_2017,musso_HowDoesCosmic_2018,song_HaloMassQuenching_2021}.
In future work, we intend to test these models by splicing a galaxy at different distances from a cosmic filament.

The gene-splicing method may also prove useful in the study of the secondary bias problem \citep{gao_AssemblyBiasClustering_2007,dalal_HaloAssemblyBias_2008,hahn_TidalEffectsEnvironment_2009}.
In particular, it enables direct tests of how the anisotropy in the environment affects the relationship between bias and concentration \citep{paranjape_HaloAssemblyBias_2018}.
In this paper, we have applied the splicing operation to exactly fix the density field in a \emph{finite} region of space.
This however does not mean that other fields, such as the velocity field or the tidal shear, are fixed as well in the region.
In order to fix the initial shear, we could apply the gene-splicing operation to the potential rather than the density field. This would still fix the density field in the spliced region but also the tidal shear, since both can be computed from local derivatives of the potential.
However, since the potential is a smoother field (\ie{} has longer-range correlations) than the density, the gene-splicing operation will be less localised in this alternative formulation. Consequently, while the large scale structure surrounding the spliced region is minimally affected when we splice overdensity (as in the present work), it may be more strongly changed when splicing potential.
This extension to splicing potential rather than density will be explored in future work.

\section*{Acknowledgements}

CC thanks S.~Codis and M.~Musso for stimulating discussions.
LLS thanks E.~Komatsu for useful comments on the manuscript.
This project has received funding from the European Union's Horizon 2020 research and innovation programme under grant agreement No.\ 818085 GMGalaxies.
HVP's work was partially supported by the research project grant `Understanding the Dynamic Universe' funded by the Knut and Alice Wallenberg Foundation under Dnr KAW 2018.0067.
AP was supported by the Royal Society.
{This work used computing equipment funded by the Research Capital Investment Fund (RCIF) provided by UKRI, and partially funded by the UCL Cosmoparticle Initiative.}
The analysis was carried out using
\textsc{Colossus} \citep{diemer_COLOSSUSPythonToolkit_2018},
\textsc{Jupyter} notebooks \citep{soton403913},
\textsc{Matplotlib} \citep{hunter2007matplotlib},
\textsc{Numpy} \citep{harris_ArrayProgrammingNumPy_2020},
\textsc{Pynbody} \citep{pontzen_PynbodyNBodySPH_2013},
\textsc{Python},
\textsc{Tangos} \citep{pontzen_TANGOSAgileNumerical_2018} and
\textsc{Yt} \citep{turk_YtMulticodeAnalysis_2011}.

\section*{Author contributions}

The main roles of the authors were, using the CRediT (Contribution Roles Taxonomy) system (\url{https://authorservices.wiley.com/author-resources/Journal-Authors/open-access/credit.html}):

{\bf CC:} conceptualisation; methodology; validation; investigation; data curation; formal analysis; writing -- original draft; visualisation.

{\bf AP:} conceptualisation; methodology; software; validation and interpretation; writing -- review and editing; funding acquisition.

{\bf HVP:} conceptualisation; validation and interpretation; writing -- review and editing.

{\bf LLS:} conceptualisation; validation and interpretation; writing -- review and editing.

\section*{Data availability}
The data underlying this article will be shared on reasonable request to the corresponding author.

\bibliographystyle{mnras}
\bibliography{authors}

\appendix
\section{Derivation of the splicing technique}%
\label{sec:splicing-mathematical-derivation}

In this Appendix, we present an outline of the technique by which a given initial density field can be `spliced' into another while remaining consistent with a $\Lambda$CDM power spectrum.

We assume the initial conditions to be in the linear regime so that we can write any density field as
\begin{equation}
    \rho(\vec{x}) = \rho_0 (1 + \delta(\vec{x})),
\end{equation}
where $\delta(\vec{x})$ is a Gaussian random field.
The problem of generating initial conditions in the case of numerical simulations involves a list of $N$ discretised cells located at position $\vec{x}_i$.
In order to draw ICs, we need to draw $N$ values from a multivariate normal distribution with $N$ dimensions with mean value $\vec{\mu}$ and covariance $\mat{\Cov}$.
For $\Lambda$CDM initial conditions, $\vec{\mu}=0$, while the effect of $\mat{\Cov}$ on any vector can be computed by performing a discrete Fourier transform, multiplying by the power spectrum, and transforming back to real space. The effect of $\mat{\Cov}^{-1}$ is obtained through a similar sequence, dividing by the power spectrum instead of multiplying.
Note that the splicing method is not limited to the density field and can be applied to any Gaussian random field, such as the gravitational potential field.

Extending the description of \cref{sec:qualitative-presentation-splicing}, let us write the  discretised independent random fields as $\vec{a}$ and $\vec{b}$ respectively. We also introduce a mask matrix $\mat{M}$ which zeros pixels lying outside the selected region $\Gamma$, while leaving those inside the region untouched. We will also use the shorthand $\overline{\mat{M}} \equiv \mathbb{I} - \mat{M}$; functionally, $\overline{\mat{M}}$ zeros all pixels inside the mask, while retaining the value of those outside.

A spliced field $\vec{f}$ satisfies the defining relation:
\begin{align}
    \text{minimise } Q &= {(\vec{b} - \vec{f})}^\dagger \mat{\Cov}^{-1} (\vec{b} - \vec{f}), \nonumber \\
    \text{subject to } \mat{M} \vec{f} &=  \mat{M}\vec{a}. \label{eq:splicing-minimisation}
\end{align}
To solve this constrained quadratic minimisation, we split the problem into two systems -- one of which is fully constrained, while the other is entirely unconstrained. Specifically, we write
\begin{equation}
    \vec{\Delta} = \vec{f} - \vec{b},
    \label{eq:Delta_def}
\end{equation}
with which definition, the constraint becomes $\mat{M} \vec{\Delta} = \mat{M}(\vec{a} - \vec{b})$.
The solution for $\vec{\Delta}$ in the system~\eqref{eq:splicing-minimisation} then takes the form
\begin{equation}
    \vec{\Delta} = \mat{M}(\vec{a} - \vec{b}) + \overline{\mat{M}} \vec{\alpha}.
    \label{eq:Delta_alpha}
\end{equation}
The constraint is satisfied by construction, while the vector $\vec{\alpha}$ is defined implicitly by the minimisation of $Q$.
Back-substituting our definitions into $Q$, we have
\begin{equation}
    Q = \vec{\alpha}^{\dagger} \overline{\mat{M}} \mat{\Cov}^{-1} \overline{\mat{M}} \vec{\alpha} + \vec{\alpha}^{\dagger} \overline{\mat{M}} \mat{\Cov}^{-1} \mat{M} (\vec{a} - \vec{b}) + \mathrm{c.c.} + \mathrm{const.}
\end{equation}
Here c.c.\ indicates the complex conjugate of the preceding term, while the
constant term does not depend on the quantity we are now optimising, $\vec{\alpha}$,
and therefore does not need explicit calculation. Here we have used the fact that all the matrices
$\mat{\Cov}$, $\mat{M}$ and $\overline{\mat{M}}$ are Hermitian.

Minimising $Q$ (now without any constraints) requires
\begin{equation}
    \overline{\mat{M}} \mat{\Cov}^{-1} \overline{\mat{M}}\vec{\alpha} = \overline{\mat{M}} \mat{\Cov}^{-1} \mat{M} (\vec{b} - \vec{a}).
\end{equation}
This is an equation of the form $\mat{A} \vec{\alpha} = \vec{z}$ with $\mat{A} = \overline{\mat{M}} \mat{\Cov}^{-1} \overline{\mat{M}}$ and $\vec{z} = \overline{\mat{M}} \mat{\Cov}^{-1} \mat{M} (\vec{b} - \vec{a})$, which can be solved by standard conjugate gradient optimisation methods. At each step of the optimisation, a discrete Fourier transform and its inverse will be computed in order to multiply by $\mat{\Cov}^{-1}$.

Other than its null space, $\mat{A}$ has a similar spectrum to $\mat{\Cov}^{-1}$, and therefore we improve the convergence rate of the conjugate gradient method by pre-conditioning the problem with the matrix $\mat{\Cov}$.
Once $\vec{\alpha}$ is obtained, we can back-substitute it in \cref{eq:Delta_def,eq:Delta_alpha} to find $\vec{f}$.
The ICs are generated by finally computing particle displacements from the density field.

In our implementation, we stop the conjugate gradient iterations once the Euclidian norm of the residuals becomes smaller than $\|\vec{z}\|/\num{e6}$.
The convergence is typically achieved in a few steps in 1D, and a few hundred steps in 3D depending on the size and shape of the spliced region, and the size of the grid it is spliced into.
We find that the time-to-solution scales roughly as $N^4$, where $N$ is the number of cells in each direction. For the simulations in this work, we obtained solutions to the splicing minimisation within four minutes on \num{32} cores; given the scaling is barely worse than for a single FFT which scales as $N^3 \ln N$, the method can be applied to much larger simulations.

\bsp{}  %
\label{lastpage}
\end{document}